%% file: main.tex
\documentclass[runningheads]{llncs}
\usepackage[hidelinks]{hyperref}
\usepackage{todonotes}
\usepackage[T1]{fontenc}
\usepackage{graphicx}

\usepackage{algorithm}
\usepackage{algpseudocode}

\begin{document}
\title{The ParlaSpeech Collection of Automatically Generated Speech and Text Datasets from Parliamentary Proceedings}
\titlerunning{The ParlaSpeech Collection of Speech and Text Datasets}
%
\author{Nikola Ljube\v{s}i\'{c}\inst{1,2}\orcidID{0000-0001-7169-9152} \and
Peter Rupnik \inst{1}\orcidID{0009-0000-9700-3686} \and
Danijel Kor\v{z}inek \inst{3}\orcidID{0000-0002-2916-4856}}
\authorrunning{N. Ljube\v{s}i\'{c} et al.}
%
\institute{Jožef Stefan Institute, Ljubljana, Slovenia \\
    \email{nikola.ljubesic@ijs.si}\\
    \email{peter.rupnik@ijs.si}\\
    \and
    Faculty of Computer and Information Science, University of Ljubljana, Slovenia
    \and
    Polish-Japanese Academy of Information Technology, Warsaw, Poland\\
    \email{danijel@pjwstk.edu.pl}}
\maketitle              
\begin{abstract}
    Recent significant improvements in speech and language technologies come both from self-supervised approaches over raw language data as well as various types of explicit supervision. To ensure high-quality processing of spoken data, the most useful type of explicit supervision is still the alignment between the speech signal and its corresponding text transcript, which is a data type that is not available for many languages. In this paper, we present our approach to building large and open speech-and-text-aligned datasets of less-resourced languages based on transcripts of parliamentary proceedings and their recordings. Our starting point are the ParlaMint comparable corpora of transcripts of parliamentary proceedings of 26 national European parliaments. In the pilot run on expanding the ParlaMint corpora with aligned publicly available recordings, we focus on three Slavic languages, namely Croatian, Polish, and Serbian. The main challenge of our approach is the lack of any global alignment between the ParlaMint texts and the available recordings, as well as the sometimes varying data order in each of the modalities, which requires a novel approach in aligning long sequences of text and audio in a large search space. The results of this pilot run are three high-quality datasets that span more than 5,000 hours of speech and accompanying text transcripts. Although these datasets already make a huge difference in the availability of spoken and textual data for the three languages, we want to emphasize the potential of the presented approach in building similar datasets for many more languages.
\end{abstract}

\vspace{10cm}

\section{Introduction}

Although self-supervision has been shown to be the main driving force in recent drastic improvements in intelligent data processing of different data modalities, including image~\cite{rani2023self}, video~\cite{schiappa2023self}, text~\cite{achiam2023gpt}, speech~\cite{baevski2020wav2vec}, as well as different modalities~\cite{baevski2022data2vec} explicit supervision via text and speech correspondence has proven to still be the most valuable signal in developing technologies that allow processing of spoken data~\cite{radford2023robust}. In this paper, we are tackling one of the more promising approaches to obtain text and speech aligned data for a large number of languages, namely through parliamentary recordings and their available manual transcripts.

\subsection{Motivation}

The availability of speech and text datasets differs drastically between languages, the Common Voice project~\cite{ardila2019common} being a good approximation of the overall language distribution among such data: a few languages having very good coverage, some languages having decent coverage, and a long tail of languages with very limited or no coverage. The three pilot languages that we are dealing with in this paper depict the problem of the long tail very clearly. Polish, an official EU language with more than 40 million speakers, has 180 hours of material in the latest version of the dataset, Serbian has 12 hours, while Croatian, another official EU language with 4 million speakers, is still not present in the dataset. Croatian is not only not present in this data set, but before our efforts, there was no single open speech and text data set available for that language~\cite{ljubevsic2022parlaspeech}.

A convenient source of speech and the corresponding text data for official languages are parliamentary proceedings. This is because of regulations that often require the transcripts of parliamentary proceedings to be publicly available, as well as because of the frequent availability of the recordings of the proceedings as part of the public domain. The availability of speech data in the public domain is especially useful as it resolves quite a number of questions related to the biometric properties of the speech signal and the underlying privacy issues.


\subsection{Prerequisites}

In recent years, two iterations of the ParlaMint project~\cite{erjavec2023parlamint} were funded by the CLARIN ERIC infrastructure on language resources and technologies. The main goal was to uniformly encode transcripts of parliamentary proceedings of various European parliaments. With these efforts, the availability of parliamentary transcripts for almost all official European languages has improved drastically. The current number of national parliaments covered is 26.

As part of the third iteration of the ParlaMint project, a pilot, coined ParlaSpeech, has been run with the goal of exploiting the improved availability of the textual transcripts by aligning these transcripts to the recordings of parliamentary sessions, ensuring the availability of text and speech datasets in languages not previously adequately covered with such data. For this pilot, the three already mentioned languages were chosen: Croatian, Polish and Serbian. The reasons for including exactly these three languages in the pilot were: (1)~there is little to no data available for these languages, (2)~there is a significant amount of transcript data available inside ParlaMint, (3)~the main proponents of the ParlaSpeech pilot have good knowledge of both the ParlaMint datasets for these languages, as well as knowledge of the languages themselves, (4)~the recordings from these parliaments are available through YouTube.

Before scaling to all three languages, an initial run of ParlaSpeech was performed only for the Croatian language~\cite{ljubevsic2022parlaspeech}, resulting in the first publicly available text and speech dataset for the Croatian language of 1,816 hours in size, together with the first ASR systems trained on a subset of available data. In this paper, we describe the second iteration of this effort, where we took the lessons learned from the initial iteration and scaled it up to three languages, with the obvious goal of scaling the approach further to even more languages in follow-up activities.

\subsection{Main Challenges}

While dealing with the problem of aligning parliamentary transcripts to recordings of parliamentary sessions, the following main challenges were identified: (1) parts of audio recordings are not transcribed, (2) some transcribed recordings are not released to the public, (3) parts of the recordings are transcribed with significant deviations from what has actually been said, (4) the metadata released with the recordings and the transcripts, such as the date of the session being recorded or transcribed, do not correspond, and (5) the order of texts in the transcripts does not follow the order of the events in the recording.

\subsection{Similar Projects}

Exploiting parliamentary data to build spoken corpora or text and speech datasets is by far not a new idea. There have been successful efforts in building such open datasets for Swiss French and German~\cite{imseng2012mediaparl}, Icelandic~\cite{helgadottir2017building}, Danish~\cite{kirkedal2020ft}, Czech~\cite{kratochvil2020large,kopp2021parczech}, Swiss German~\cite{pluss2022sds} Norwegian~\cite{solberg2022norwegian}, and Finnish~\cite{virkkunen2023finnish}. However, this is the first project where an approach that can be scaled to many languages is developed. A crucial component of this approach is, of course, the availability of comparable text transcripts from the ParlaMint project.

\subsection{Paper Overview}

The remainder of the paper is structured as follows. In Section 2 the problem of matching long sequences of text and speech is described, especially in light of parliamentary data and its challenges. Section 3 describes the proposed alignment procedure. Section 4 discusses post-processing decisions motivated by the release of each dataset in three flavors, described in Section 5, namely as (1)~a FAIR repository entry, (2)~a HuggingFace dataset for simplifying usage for automatic speech recognition and related tasks, and (3)~a corpus in a concordancer enabling advanced search through the dataset. The paper ends with a conclusion and a description of future directions.

\section{Long Speech to Text Sequence Alignment}

The core of the problem discussed in the paper is aligning a long audio recording of speech to a long body of text to acquire word-level timestamps. The audio is derived from a video archive available online, and the text is a corpus derived from a human transcript of the audio made at an unknown point, usually by a government entity in accordance with local laws and traditions. The length of the video is commonly several hours, and the text usually spans a whole day's session of parliamentary proceedings. A single session is sometimes divided into several video recordings. There is no guarantee of completeness or ordering in the two sequences: there is a possibility that some information is lacking in either sequence, and the order of chunks of tokens within the text sequence could differ from the temporal order of the audio. Furthermore, the accuracy of the transcript is not ideal, because the purpose of the stenography is to create a transcript that is easy to read, rather than something that precisely depicts the audio, with all the details specific for spoken communication, such as overlapping speech and disfluencies.

\subsection{Existing Approches}

The idea of aligning long sequences is not new. Typical forced alignment suffers from exponential growth in complexity with respect to the length of the sequence being aligned, but there are methods to overcome these limitations. In~\cite{katsamanis2011sailalign}, the approach was to first perform text-to-text alignment of the actual reference to the transcript perceived by the acoustic model. The acoustic model transcript is acquired using an ASR system fine-tuned to the real reference, but allowing for discrepancies through the use of an N-gram language model. The text-to-text alignment finds regions of exact matches (i.e., speech landmarks) with gaps that do not match. The matches are assumed to be correct and the mismatches are recursively re-aligned, using the same procedure, until convergence. A similar approach was used in~\cite{panayotov2015librispeech} to create an alignment between audiobooks of the LibriVox project and their original text present in Project Gutenberg. That procedure was a bit more complex, as it included a more advanced ASR solution, a better language model adaptation technique, and confidence-based filtering of ASR output. The general idea remained the same.

\subsection{Our Approach}

\begin{figure}
    \centering
    \includegraphics[width=\linewidth]{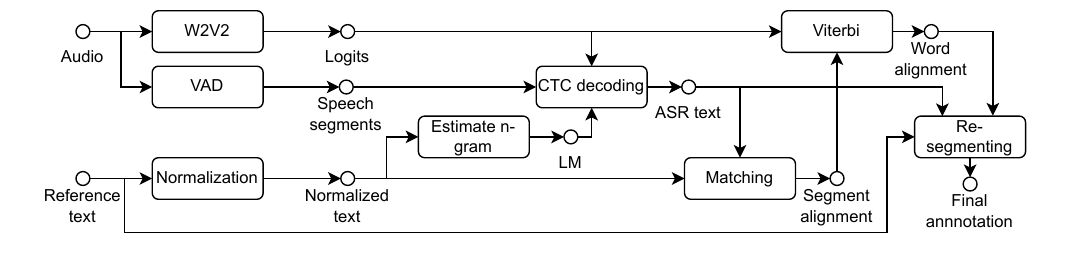}
    \caption{Diagram of long speech to text sequence alignment pipeline for processing a single audio-text file pair. Circles are intermediary data structures. Rectangles are processes.}
    \label{fig:pipeline}
\end{figure}

This paper describes a method that is an adaptation of the above approaches with two major upgrades: (1) the utilization of a more modern end-to-end speech recognition system and (2) modifications of the text-to-text matching routine to suit the requirements of the data and the purpose of the final product. More specifically, the purpose of the older method described above is to acquire the most accurate alignment while assuming the input data to be complete and accurate. The latter method, on the other hand, simply looks for the creation of a decent corpus used to train speech recognition models - completeness of the final product is of lesser concern. In the case of this paper, we know the data is incomplete, but we strive to achieve as much coverage as possible because the purpose is to index the data and allow further research in various settings (e.g., linguistic studies or political science studies). This is reflected in the heuristics described below. Following is a description of the complete pipeline and all its components as illustrated in Figure~\ref{fig:pipeline}.

\section{Processing pipeline description}

The procedure for aligning a large dataset begins with a collection of audio files and a long text corpus. We iterate by audio files and need to find a chunk of text that is reasonably long to cover the contents of that audio recording. The corpus is divided into sections with corresponding metadata, but in our experience, there is frequently a mismatch between the recording and transcript metadata. That is why we utilize a statistical analysis - ratio of n-gram coverage, comparing the reference transcripts to ASR outputs of each file to see what is the best match. Sometimes, several transcripts are matched to a single recording, and vice versa. For the remainder of this section, we will assume that the matching between the recordings and transcript is present and start the pipeline description from a single audio recording and transcript covering the contents of that file.

\subsection{Audio processing}

The pipeline for processing a single recording and transcription pair is designed to facilitate mass data processing, so the intermediary results are saved in a cache for reuse in multiple stages along the way. Such is the case with speech processing -- instead of performing ASR in one go, we first compute the Wav2Vec2-XLS-R (W2V2) model \cite{conneau2020xslr} logarithmic likelihoods (logits) into a file and then use it both for ASR decoding and Viterbi alignment, which occurs later in the process.

Except for the calculation of W2V2 logits, we also use Voice Activity Detection (VAD) to figure out which parts of the recordings contain speech that is likely being transcribed by the human transcribers. We use the pyannote~\cite{Bredin23} package to extract speech segments and then remove segments with an energy level below -45 dB RMS. This is to remove most of the background conversations that are generally not transcribed. Both W2V2 logits extraction and VAD processing are computed on the GPU and caching their outputs allows for optimal use of the hardware. All other processes are computed on the CPU.

\subsection{Text Pre-processing}

At the same time, we normalize the reference text. Depending on the W2V2 model, the output of ASR may or may not contain digits, but usually does not contain most of the symbols, punctuation or capitalization. To better accommodate the matching of the ASR output to the human transcript, we need to normalize the human transcript to remove any punctuation and capitalization, as well as convert any symbolic text into its pronounced form. This is a somewhat language-dependent procedure, and although it is well researched~\cite{bakhturina2022normalization}, we had to rely on custom rule-based solutions for the languages and corpora being prepared in this paper. Even though it is a common procedure in, e.g., text-to-speech software, no high-quality open-source solutions existed for the languages analyzed at the time of performing the research.

\subsection{Language Modeling and Speech Recognition}

The normalized text is also used to prepare the language model (LM) for the ASR decoding phase of the procedure. We tried to train the model only on the text in the transcript being processed, but we obtained much better results by combining all the transcripts and creating a single LM for all the files being processed. This is most likely due to the better statistics acquired with a larger quantity of text. We used the SRILM toolkit~\cite{stolcke2002srilm} to train a Knesser-Ney discounted 3-gram model with interpolation. We then use the pyctcdecode\footnote{\url{https://github.com/kensho-technologies/pyctcdecode}} package to generate the ASR output text based on the W2V2 logits and the language model. The VAD output is also used at this stage to determine which parts of the audio should be processed.

\subsection{Matching of Automatic to Reference text}

\begin{figure}
    \centering
    \includegraphics[width=\linewidth]{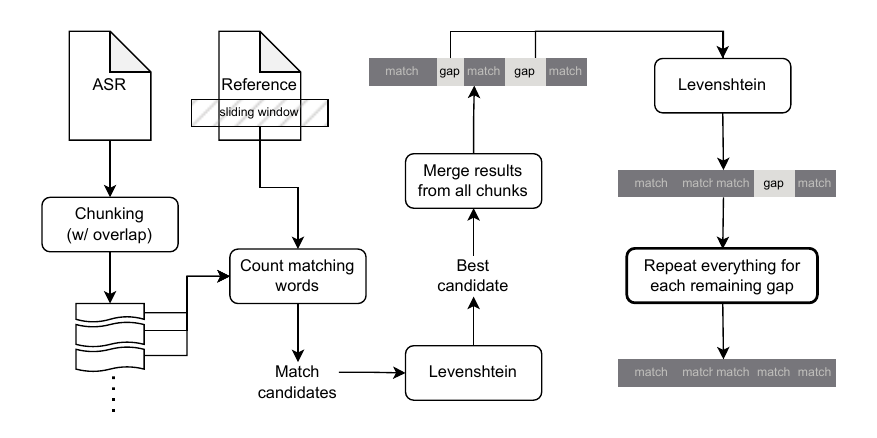}
    \caption{Illustration of the matching algorithm. The purpose is to find portions of the reference that match the ASR output. Sequence and accuracy is not guaranteed.}
    \label{fig:matching}
\end{figure}

The next step in the pipeline is to match the generated ASR output with the normalized reference text. The procedure, as illustrated in Figure~\ref{fig:matching}, starts by looking for potential matches between the two text sequences using a word histogram of a sliding window. Next, each match is evaluated using the Levenshtein distance to find the best match. The nonmatching prefix and suffix (so any insertion or deletion at the start and end of sequence) is rejected, and everything else is treated as the final match. Depending on the chosen thresholds, this method leaves gaps in the final result. We then try to force the matches in those gaps by using the Levenshtein comparison again. Sometimes this still leaves gaps, particularly if they are large or if the order of the sequences does not match. In that case, we try and repeat the whole procedure outlined above, starting with the histogram-based search of the whole reference, but only within the remaining ASR gaps. The statistics of coverage (i.e. how many words are matched in either sequence) are computed along each phase of the procedure, which helps in tuning the thresholds and other hyperparameters.

The output of the matching is a list of audio segments and their matching reference text. These segments can span many words, so in order to obtain time offsets for each individual word, we need to re-align the audio to the actual reference text rather than the ASR output obtained earlier. For this, we use a simple Viterbi forced alignment algorithm~\cite{Rabiner1989ATO} to match the character sequence to the output of the W2V2 model. One feature of the W2V2 model is the presence of word delimiter tokens in the output: to obtain word level offsets, it is sufficient to look for the location of the word delimiters within the aligned sequence.

\subsection{Post-processing}

In the final step of the pipeline, all the information from previous stages is combined to create one coherent, aligned annotation of the file. We use a mapping between the original and normalized reference text to project the time offsets onto the original un-normalized token sequence. We also add the ASR output, temporally aligned with the reference sequence above. This is both for visualizing possible errors in human transcripts (by calculating the word error rate between the ASR output and the reference text) and for providing automatic transcription for parts of audio that human transcribers did not transcribe. Finally, we map everything into chunks that were used in the original reference text corpus. Each chunk of the original corpus contains a unique ID, which allows us to combine our acoustic time annotation with other forms of annotation and metadata present in the original text corpus.

\section{Segmentation and Filtering}

In this section, we describe the further segmentation and accompanying filtering from our text and audio alignment process, presented in the previous section. This segmentation and filtering are necessary to generate the datasets and corpora that we are currently considering to be most useful for downstream usage. We describe such three downstream cases in the next section.


The output of the alignment process from the previous section are JSON files, one per each original audio file, consisting of three types of entries: (1)~an ASR transcription of the part of the audio file that could not be matched to any part of the ParlaMint corpus,  (2)~a speech transcript from the ParlaMint corpus that could not have been aligned with this part of this audio file, but its surrounding transcripts could, and (3)~the speech transcript from the ParlaMint corpus together with the ASR transcription that was matched to that particular ParlaMint transcript, along with a list of predicted word alignments consisting of character offsets (referencing to the ParlaMint speech transcript) and millisecond offsets (referencing to the audio recording).

To obtain streamlined datasets from this matching output, which would be better suited for downstream applications such as aligned text and speech datasets for training automatic speech recognition, or spoken corpora for linguistic purposes, a series of additional filterings and segmentations have been performed.

The first filtering iteration was performed at the level of speeches, removing all ParlaMint speech transcripts without audio alignment, as well as speeches with alignment, but with an estimated character error rate between the ParlaMint transcript and the aligned ASR transcription higher than or equal to 60\%. The aim of this filter was to remove nonaligned portions of the data, as well as those portions where even partial alignments are very questionable.

In the next step, each speech transcript was segmented into sentences to filter out parts of the speeches with a lack of correspondence. For each sentence covered by word alignment information, the character error rate was calculated again between the ParlaMint transcript and the ASR transcription, and all sentences with a character error rate greater than 10\% were filtered out. With this filter, sentences with deviation between the spoken signal and the transcript have been discarded.

A final filtering at the sentence level was performed in cases where the ratio of the length of the audio in milliseconds and the length of the transcript in characters were greater than $0.2$. Namely, in some cases, the alignment process matched a sentence to part of the audio with longer or shorter breaks in the work of the parliament, making the audio unrealistically longer than what would be expected given the length of the transcript. Given that these breaks are mostly muted audio, such imperfections could not have been filtered out with the previous filters based on the character error rates.

All three filtering thresholds were defined by inspecting samples of data and manually identifying reasonable cut-off points.

To estimate the yield rate of the whole matching and filtering procedure, we calculated the percentage of ASR transcriptions that were successfully matched to the ParlaMint transcripts after all filterings. We performed this yield estimation on the Croatian data as we are rather positive that the recordings are to the most part covered with the ParlaMint transcripts. The yield rate for matching the available audio information with the textual transcript in this particular case was 74\%. Given our experience with the data, which also includes manual analyses of the non-aligned ASR transcriptions, the main reasons, in order of prevalence, for parts of the spoken content not being aligned to the ParlaMint transcripts are: (1) speech not being transcribed within the parliament, (2) transcripts differing from the spoken word, (3) ASR errors, and (4) matching errors.

\section{The Dataset Releases} 

In this section, we describe the three encodings of our datasets aimed at the specific downstream use cases: master CLARIN.SI FAIR (findable, accessible, interoperable, reusable) repository entries aimed at archiving all available information, HuggingFace datasets practical for using our data on tasks such as training automatic speech recognition and various speech classification models, and linguistically annotated corpora that allow complex linguistically informed searches through the datasets.

\subsection{FAIR Repository Entries}

As the master release of the produced datasets we have prepared jsonl files, each line covering a single sentence from the ParlaMint corpus, with a reference to the corresponding flac audio file, and word-level alignment containing character offset and millisecond offset information.

In addition to the spoken and textual content and their alignment information, a significant amount of metadata present in the ParlaMint corpus has also been included in these datasets. Inter alia, the following information on the speaker was included: the role of the speaker (are they speaking as a chairperson or an MP), the party they belong to, the political orientation of the party, whether the party is in coalition or opposition at the time of the speech, and the gender and their year of birth of the speaker.

We publish such prepared datasets on the FAIR (findable, accessible, interoperable, reusable) repository of CLARIN.SI,\footnote{\url{https://www.clarin.si/repository/xmlui/}}, the Slovenian national node of the CLARIN ERIC infrastructure on language resources and technologies.\footnote{\url{https://www.clarin.eu}}

The statistics on the size of each of these releases are presented in Table~\ref{tab:stats}. From the presented numbers, it is obvious that the Croatian dataset is by far the largest. While for both Croatian and Polish, all available data were processed, for the Serbian dataset, only 1000 audio files out of almost 4500 files have been processed by now. We hope that we will be able to further expand the Serbian dataset in one of the next ParlaSpeech data collection releases.

\begin{table}[]
    \centering
    \input{code/stats.tex}
    \vspace{0.3cm}
    \caption{The statistics on the size of the three ParlaSpeech datasets}
    \label{tab:stats}
\end{table}
\begin{center}

\end{center}

The repository entries can be accessed through persistent identifiers for Croatian\footnote{\url{http://hdl.handle.net/11356/1914}}, Polish\footnote{\url{http://hdl.handle.net/11356/1686}} and Serbian\footnote{\url{http://hdl.handle.net/11356/1834}}.

\subsection{HuggingFace Datasets}

The second release of the dataset is through the HuggingFace Datasets Hub,\footnote{\url{https://huggingface.co/datasets}}, which allows technical users to gain access to all three data sets using just a few lines of code.


The data sets are again available separately for Croatian\footnote{\url{https://huggingface.co/datasets/classla/ParlaSpeech-HR}}, Polish\footnote{\url{https://huggingface.co/datasets/classla/ParlaSpeech-PL}}, and Serbian\footnote{\url{https://huggingface.co/datasets/classla/ParlaSpeech-RS}}.

Given that parts of the available speaker metadata were included in the HuggingFace datasets, such as the gender, age, party affiliation, whether the speaker was in coalition or opposition during the speech, the data are useful for so much more than just automatic speech recognition. We are looking forward to all the interesting use cases that this data availability will produce.

\subsection{Spoken Corpora via Concordancer}

Finally, the third availability of the dataset is aimed at the use of linguists and phoneticians. Each dataset has been made available through the CLARIN.SI concordancer~\footnote{\url{https://www.clarin.si/ske/}} with the help of which the textual transcript can be searched with the Corpus Query Language, and the recording of each search result can be played back.

To enable more detailed searches using the Corpus Query Language, each of the sentences in the corpus was linguistically annotated, splitting each sentence into words and annotating each word with the part of speech, the morphosyntactic features, and the lemma of the word. For Croatian and Serbian, the CLASSLA-Stanza tool~\cite{tervcon2023classla} was used, while for Polish we applied the Stanza tool~\cite{qi2020stanza}.

To make the playback of the recordings as user-friendly as possible, with the median length of sentence recordings between 5 and 10 seconds, depending on the language, for this release we have performed another segmentation of the data with the aim of obtaining recordings in length between 3 and 6 seconds. If a researcher requires recording of the entire sentence, it can still be accessed in the metadata of each sentence.

The availability of the datasets through concordancers is again separated by language, having a separate corpus for Croatian\footnote{\url{https://www.clarin.si/ske/\#dashboard?corpname=parlaspeech\_hr}}, Polish\footnote{\url{https://www.clarin.si/ske/\#dashboard?corpname=parlaspeech\_pl}}, and Serbian~\footnote{\url{https://www.clarin.si/ske/\#dashboard?corpname=parlaspeech\_rs}}.

The intended use of the concordancer is to simplify linguists and phoneticians' identification of linguistic patterns that they are interested in, and accessing their recordings that can be further processed in specific tools such as Praat~\cite{styler2013using}, or Exmaralda~\cite{schmidt2014exmaralda}.

\begin{figure}
    \centering
    \includegraphics[width=\linewidth]{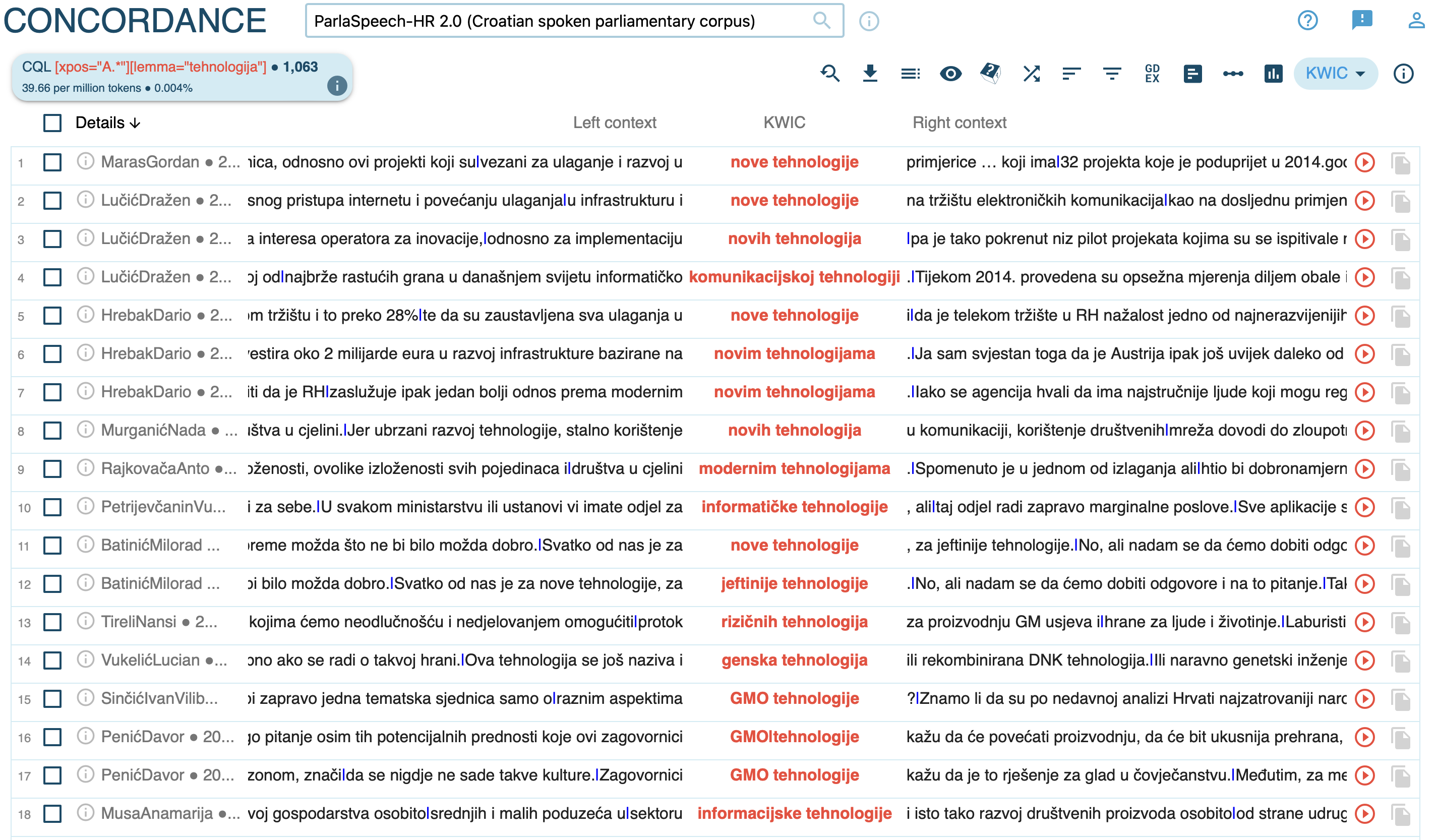}
    \caption{Example of a search result on the noun ``tehnologija'' with a preceding adjective in the concordancer of the Croatian corpus. The recording can be accessed to the right, the metadata to the left.}
    \label{fig:ske}
\end{figure}

An example of the result of a search for the noun ``tehnologija'' with a preceding adjective in the Croatian corpus is presented in Figure~\ref{fig:ske}. The sought phrase is colored red, the surrounding context is colored black, with an icon for playing the recording of the phrase to the right, and a link to the speaker and sentence metadata to the left.

\section{Conclusion}

In this paper, we have presented a robust and scalable approach to aligning thousands of hours of recordings of parliamentary proceedings with their manual transcripts released by the parliaments. This process includes a series of challenges, the biggest ones being the non-correspondence in either the data coverage or the data order in any of the two data modalities. Furthermore, deviations in the transcripts from the spoken words are a rather frequent phenomenon. The reason for these deviations is that the transcripts are not aimed at performing linguistic research or speech technology development, but to ensure availability of the parliamentary discussions for the general public.

We have described three complementary approaches to releasing the final datasets - (1)~a complete, master release through a FAIR repository to ensure maximum availability and reusability of the data, (2)~the opportunistic release through the HuggingFace Datasets Hub to simplify speech technology development, not only on the problem of automatic speech recognition, but also additional tasks such as demographic prediction due to availability of rich metadata, and (3)~the release in form of spoken corpora available through a linguistic concordancer, which allows linguists to search the transcripts, enriched with linguistic features such as part of speech, lemma, and morphosyntactic features, as well as to listen or to retrieve the recordings of their search results.

There are limitations to our work. The first is on the side of the data that come from the rather limited parliamentary domain and, even more, they are filtered by the correspondence between the recording and the transcript, which removes all the content where transcribers were performing stronger edits, such as disagreements, disfluencies, etc. The second is on the side of the method that requires an at least partially functioning speech encoder and related ASR system, but also the availability of the transcripts and the recordings of the parliamentary sessions.

With the presented results of more than 5,000 hours of corresponding speech and text in three less-resourced Slavic languages, we are of the opinion that we have just scratched the surface of what the ParlaSpeech concept can bring to the research community. We will primarily focus on adding additional languages to the ParlaSpeech collection, as there is a significant number of the current ParlaMint languages that would immensely profit from the availability of ParlaSpeech data for that language. In parallel with that, we will use the ParlaSpeech data in both technology development and linguistic and communication research, especially looking out for various types of biases in the data themselves, as well as biases that we have introduced with our matching and filtering procedures.

\begin{credits}
    \subsubsection{\ackname}
    The research presented in this paper was conducted within the research project ``ParlaMint: Towards Comparable Parliamentary Corpora'' funded by CLARIN ERIC. The research was also co-funded by the research project titled ``Basic Research for the Development of Spoken Language Resources and Speech Technologies for the Slovenian Language'' (J7-4642), and withing the research programme ``Language resources and technologies for Slovene'' (P6-0411), both funded by the Slovenian Research and Innovation Agency (ARIS).

\end{credits}

\bibliography{bibliography}
\bibliographystyle{splncs04}

\end{document}

%% file: code/stats.tex
\begin{tabular}{|l|r|r|r|}
\hline
corpus & \multicolumn{1}{c|}{HR} & \multicolumn{1}{c|}{PL} & \multicolumn{1}{c|}{RS} \\
\hline
size (GB) & 179.0 & 60.8 & 57.7 \\
duration (h) & 3110.39 & 1009.82 & 896.22 \\
sentences & 922 679 & 535 465 & 290 778 \\
words & 24 755 742 & 7 515 333 & 7 024 293 \\
characters & 150 970 948 & 52 724 103 & 42 638 259 \\
median sentence (s) & 9.62 & 4.94 & 8.74 \\
\hline
\end{tabular}

%% file: main.bbl
\begin{thebibliography}{10}
\providecommand{\url}[1]{\texttt{#1}}
\providecommand{\urlprefix}{URL }
\providecommand{\doi}[1]{https://doi.org/#1}

\bibitem{achiam2023gpt}
Achiam, J., Adler, S., Agarwal, S., Ahmad, L., Akkaya, I., Aleman, F.L., Almeida, D., Altenschmidt, J., Altman, S., Anadkat, S., et~al.: Gpt-4 technical report. arXiv preprint arXiv:2303.08774  (2023)

\bibitem{ardila2019common}
Ardila, R., Branson, M., Davis, K., Henretty, M., Kohler, M., Meyer, J., Morais, R., Saunders, L., Tyers, F.M., Weber, G.: Common voice: A massively-multilingual speech corpus. arXiv preprint arXiv:1912.06670  (2019)

\bibitem{baevski2022data2vec}
Baevski, A., Hsu, W.N., Xu, Q., Babu, A., Gu, J., Auli, M.: Data2vec: A general framework for self-supervised learning in speech, vision and language. In: International Conference on Machine Learning. pp. 1298--1312. PMLR (2022)

\bibitem{baevski2020wav2vec}
Baevski, A., Zhou, Y., Mohamed, A., Auli, M.: wav2vec 2.0: A framework for self-supervised learning of speech representations. Advances in neural information processing systems  \textbf{33},  12449--12460 (2020)

\bibitem{bakhturina2022normalization}
Bakhturina, E., Zhang, Y., Ginsburg, B.: Shallow fusion of weighted finite-state transducer and language model for text normalization (2022), \url{https://arxiv.org/abs/2203.15917}

\bibitem{Bredin23}
Bredin, H.: {pyannote.audio 2.1 speaker diarization pipeline: principle, benchmark, and recipe}. In: Proc. INTERSPEECH 2023 (2023)

\bibitem{conneau2020xslr}
Conneau, A., Baevski, A., Collobert, R., Mohamed, A., Auli, M.: Unsupervised cross-lingual representation learning for speech recognition (2020), \url{https://arxiv.org/abs/2006.13979}

\bibitem{erjavec2023parlamint}
Erjavec, T., Ogrodniczuk, M., Osenova, P., Ljube{\v{s}}i{\'c}, N., Simov, K., Pan{\v{c}}ur, A., Rudolf, M., Kopp, M., Barkarson, S., Steingr{\'\i}msson, S., et~al.: The parlamint corpora of parliamentary proceedings. Language resources and evaluation  \textbf{57}(1),  415--448 (2023)

\bibitem{helgadottir2017building}
Helgad{\'o}ttir, I.R., Kjaran, R., Nikul{\'a}sd{\'o}ttir, A.B., Gu{\dh}nason, J.: Building an asr corpus using althingi's parliamentary speeches. In: Interspeech. pp. 2163--2167 (2017)

\bibitem{imseng2012mediaparl}
Imseng, D., Bourlard, H., Caesar, H., Garner, P.N., Lecorv{\'e}, G., Nanchen, A.: Mediaparl: Bilingual mixed language accented speech database. In: 2012 IEEE spoken language technology workshop (SLT). pp. 263--268. IEEE (2012)

\bibitem{katsamanis2011sailalign}
Katsamanis, A., Black, M., Georgiou, P.G., Goldstein, L., Narayanan, S.: Sailalign: Robust long speech-text alignment. In: Proc. of workshop on new tools and methods for very-large scale phonetics research. vol.~1 (2011)

\bibitem{kirkedal2020ft}
Kirkedal, A., Stepanovi{\'c}, M., Plank, B.: Ft speech: Danish parliament speech corpus. arXiv preprint arXiv:2005.12368  (2020)

\bibitem{kopp2021parczech}
Kopp, M., Stankov, V., Kr{\u{u}}za, J.O., Stra{\v{n}}{\'{a}k}, P., Bojar, O.: Parczech 3.0: A large czech speech corpus with rich metadata. In: International Conference on Text, Speech, and Dialogue. pp. 293--304. Springer (2021)

\bibitem{kratochvil2020large}
Kratochv{\'\i}l, J., Pol{\'a}k, P., Bojar, O.: Large corpus of czech parliament plenary hearings. In: Proceedings of the Twelfth Language Resources and Evaluation Conference. pp. 6363--6367 (2020)

\bibitem{ljubevsic2022parlaspeech}
Ljube{\v{s}}i{\'c}, N., Kor{\v{z}}inek, D., Rupnik, P., Jazbec, I.P.: Parlaspeech-hr-a freely available asr dataset for croatian bootstrapped from the parlamint corpus. In: Proceedings of the workshop ParlaCLARIN III within the 13th language resources and evaluation Conference. pp. 111--116 (2022)

\bibitem{panayotov2015librispeech}
Panayotov, V., Chen, G., Povey, D., Khudanpur, S.: Librispeech: An asr corpus based on public domain audio books. In: 2015 IEEE International Conference on Acoustics, Speech and Signal Processing (ICASSP). pp. 5206--5210 (April 2015). \doi{10.1109/ICASSP.2015.7178964}, \url{https://ieeexplore.ieee.org/document/7178964}

\bibitem{pluss2022sds}
Pl{\"u}ss, M., H{\"u}rlimann, M., Cuny, M., St{\"o}ckli, A., Kapotis, N., Hartmann, J., Ulasik, M.A., Scheller, C., Schraner, Y., Jain, A., et~al.: Sds-200: A swiss german speech to standard german text corpus. arXiv preprint arXiv:2205.09501  (2022)

\bibitem{qi2020stanza}
Qi, P., Zhang, Y., Zhang, Y., Bolton, J., Manning, C.D.: Stanza: A python natural language processing toolkit for many human languages. arXiv preprint arXiv:2003.07082  (2020)

\bibitem{Rabiner1989ATO}
Rabiner, L.R.: A tutorial on hidden markov models and selected applications in speech recognition. Proc. IEEE  \textbf{77},  257--286 (1989), \url{https://api.semanticscholar.org/CorpusID:13618539}

\bibitem{radford2023robust}
Radford, A., Kim, J.W., Xu, T., Brockman, G., McLeavey, C., Sutskever, I.: Robust speech recognition via large-scale weak supervision. In: International conference on machine learning. pp. 28492--28518. PMLR (2023)

\bibitem{rani2023self}
Rani, V., Nabi, S.T., Kumar, M., Mittal, A., Kumar, K.: Self-supervised learning: A succinct review. Archives of Computational Methods in Engineering  \textbf{30}(4),  2761--2775 (2023)

\bibitem{schiappa2023self}
Schiappa, M.C., Rawat, Y.S., Shah, M.: Self-supervised learning for videos: A survey. ACM Computing Surveys  \textbf{55}(13s),  1--37 (2023)

\bibitem{schmidt2014exmaralda}
Schmidt, T., W{\"o}rner, K.: Exmaralda  (2014)

\bibitem{solberg2022norwegian}
Solberg, P.E., Ortiz, P.: The norwegian parliamentary speech corpus. arXiv preprint arXiv:2201.10881  (2022)

\bibitem{stolcke2002srilm}
Stolcke, A., et~al.: Srilm-an extensible language modeling toolkit. In: Interspeech. vol.~2002, p.~2002 (2002)

\bibitem{styler2013using}
Styler, W.: Using praat for linguistic research. University of Colorado at Boulder Phonetics Lab  (2013)

\bibitem{tervcon2023classla}
Ter{\v{c}}on, L., Ljube{\v{s}}i{\'c}, N.: Classla-stanza: The next step for linguistic processing of south slavic languages. arXiv preprint arXiv:2308.04255  (2023)

\bibitem{virkkunen2023finnish}
Virkkunen, A., Rouhe, A., Phan, N., Kurimo, M.: Finnish parliament asr corpus: Analysis, benchmarks and statistics. Language Resources and Evaluation  \textbf{57}(4),  1645--1670 (2023)

\end{thebibliography}
